\documentclass[twocolumn,apl,showpacs,preprintnumbers,amsmath,amssymb,superscriptaddress]{revtex4-1}[11pt]
\usepackage{soul}
 
\usepackage[pdftex]{graphicx}
\usepackage{amsmath}
\usepackage{amsfonts}
\usepackage{amssymb}
\usepackage{color}
\usepackage{float}
\usepackage{color}

\begin{document}
\title{Programming current reduction via enhanced asymmetry-induced thermoelectric effects in vertical nanopillar phase change memory cells}
\author{Jyotsna Bahl}
\affiliation{Centre for Research in Nanotechnology and Science, Indian Institute of Technology Bombay, Powai, Mumbai-400076, India}
\author{Bipin Rajendran}
\affiliation{Department of Electrical Engineering, Indian Institute of Technology Bombay, Powai, Mumbai-400076, India}
\author{Bhaskaran Muralidharan}
\affiliation{Department of Electrical Engineering, Indian Institute of Technology Bombay, Powai, Mumbai-400076, India}
\date{\today}
\medskip
\widetext
\begin{abstract}
Thermoelectric effects are envisioned to reduce programming currents in nanopillar phase change memory cells. However, due to the inherent symmetry in such a structure, the contribution due to thermoelectric effects on programming currents is minimal. In this work, we propose a hybrid phase change memory structure which incorporates a two-fold asymmetry specifically aimed to favorably enhance thermoelectric effects. The first asymmetry is introduced via an interface layer of low thermal conductivity and high negative Seebeck coefficient, such as, polycrystalline SiGe, between the bottom electrode contact and the active region comprising the phase change material. This results in an enhanced Peltier heating of the active material. The second one is introduced structurally via a taper that results in an angle dependent Thomson heating within the active region. Various device geometries are analyzed using 2D-axis-symmetric simulations to predict the effect on programming currents as well as for different thicknesses of the interface layer. A programming current reduction of up to $60\%$ is predicted for specific cell geometries. Remarkably, we find that due to an interplay of Thomson cooling in the electrode and the asymmetric heating profile inside the active region, the predicted programming current reduction is resilient to fabrication variability.
\end{abstract}
\maketitle
\section{Introduction} 

Phase change memories (PCMs) are non-volatile memories which use the {\it{phase change}} behavior of chalcogenide materials, typically $Ge_2Sb_2Te_5$ (GST). The phase change behavior is utilized to reversibly toggle between crystalline and amorphous phases, both having very different resistivities \cite{raoux, bipin, Burr, Wong}. A typical write operation involves the switching between a highly resistive amorphous state, called the RESET state, and a low resistance crystalline state, termed as the SET state. The conversion of SET state to RESET state is the most power expensive step associated with large programming currents. Thus, the principal aim of PCM device design is to engineer a reduction in programming currents. 
\begin{figure} [!t]
	\centering
		\includegraphics[width=3.6in,height=1.8in]{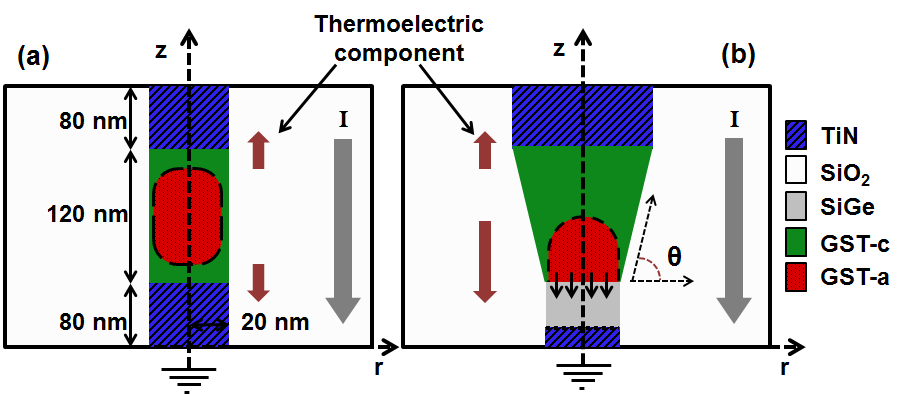}
		\caption{Device Schematics: (a) Cross sectional view of a typical vertical nanopillar PCM Cell used for the simulation. A vertical cell with an aspect ratio $3:1$ of GST height and diameter is used with TiN as the material for the top electrode. Amorphous GST region created at RESET is shown in red. With bottom and top electrodes of the same material, TiN, the symmetric vertical nanopillar cell shows minimum thermoelectric effects resulting in the amorphous region situated at the center. (b) The proposed hybrid structure: A negative Seebeck coefficient material such as polycrystalline SiGe is used as an interface layer above the bottom electrode material along with the tapered GST nanopillar to enhance the asymmetry. It is to be demonstrated here that this combination will result in increased thermoelectric effects leading to a shift of the amorphous region towards the bottom electrode and GST interface.}
		\label{Fig1}
		\end{figure} 

Design approaches to reduce programming currents are primarily geometry based \cite{bipin,Pillar,Pore,Cho,Mu_Trench,Line,Bridge,Russo}. Apart from the typical mushroom cell geometry \cite{bipin}, the most common geometry variant is the vertical pillar structure \cite{Pillar,Pop,Liu}, which may be typically fabricated using nanowires. Other methods to reduce programming currents include changes in electrode materials \cite{BEC_via}, doping of the active region \cite{Nitrogen}, and the possible use of thermoelectric effects \cite{Goodson,Azer}. The object of this work is to propose a new PCM cell design which features a geometry hybrid of the mushroom and vertical pillar geometry, in conjunction with electrode material design to explicitly utilize thermoelectric effects for programming current reduction.

A typical PCM operation involves very large internal temperature gradients and high current densities ($\sim10^{8}\,$A/cm$^2$) giving rise to electro-thermal coupling or thermoelectric effects. The Peltier effect in typical mushroom structures was accidentally observed by Suh et al., \cite{Suh} as different electrodes having different Seebeck coefficients altered the performance of the cell. On the other hand, Castro et. al., \cite{Castro} established the role of Thomson effect from a shift in the amorphous region with bias polarity. It is also well established in typical mushroom PCM cells that thermoelectric effects may favorably aid the cell performance via a reduction in programming currents \cite{Goodson,Azer}.

Vertical pillar structures \cite{Pillar,Pop,Liu}, on the one hand, are reported to have smaller programming currents due to confined size dimensions. These vertical cells also feature a less abrupt transition between the low and the high resistance states, thus making it possible to also engineer multi-bit operations \cite{Nirschl}. However, due to the symmetric nature of the vertical pillar structure \cite{Pop}, shown schematically in Fig~\ref{Fig1}(a), thermoelectric effects are not so evident. 

Thermoelectric effects are prominent when an internal temperature gradient may generate Thomson heating/cooling or an abrupt interface between different materials may generate Peltier heating/cooling. Unlike mushroom cells where the amorphous region gets formed at the interface of the bottom electrode and the active region, in a vertical pillar cell, the amorphous region is formed at the center of the active region as the heat generated in the GST dissipates equally between the top and the bottom electrode contacts. Hence, any additional heat generated due to the Peltier effect at the interface cannot contribute to the phase change process. Similarly, the vanishing temperature gradient at the central hot spot leads to negligible Thomson effect. 

In order to enhance thermoelectric effects, we therefore propose a hybrid structure as depicted in the schematic in Fig.~\ref{Fig1}(b), in which the hot spot can be shifted near to the bottom electrode-GST interface. Several studies \cite{Kim, Choi, Hubert, Wu, SiGe_Lee_2006, SiGe_Lee_2008} are dedicated to controlling the heat loss through the bottom electrode when amorphous region is formed near to interface. We hence propose the use of a interface layer ($\geq$ 10nm) with a negative Seebeck coefficient and low thermal conductivity to minimize the heat loss as well as increasing the temperature of active region, thereby lowering the programming current. The second design idea that features a taper is proposed keeping intact the vertical component of the geometry. Also, a tapered vertical structure helping in high resolution pattering is easier to fabricate \cite{raoux, Confined} in comparison to symmetric vertical structures.

Our simulations on various PCM cell designs based on the schematic in Fig.~\ref{Fig1}(b) conclusively infer on the impact of Peltier and Thomson effects on the programming currents in the RESET operation. A programming current reduction up to $60\%$ is predicted depending upon the cell dimensions. Also, the proposed structure is in compliance with scaling of the device as the proposed shaping of the cells helps in realizing the high aspect ratio and is expected to exhibit better reliability, which will be explained in the subsequent sections. Finally, we also demonstrate that the predicted lowering of programming currents are resilient to fabrication variability involved in creating the shaped cell. 

\section{Simulation Methodology} A schematic of the cross sectional view of a GST based vertical pillar PCM cell used in the simulations is shown in Fig.~\ref{Fig1}(a).  At the top and bottom of the pillar are metal electrode contacts referred to as the top electrode contact (TEC) and the bottom electrode contact (BEC). 

Unlike a mushroom cell where the GST layer is planar and wider than the narrow bottom electrode which acts as heater, in a vertical pillar cell, the confined self-heating GST pillar is generally of same dimension as that of the bottom electrode. Here, we have chosen a cylindrically symmetric structure with $20\,$nm radius and $120\,$nm height of GST (an aspect ratio $3:1$), with SiO$_{2}$ as the insulator material. Both bottom and top electrodes (TiN) are chosen to be $80\,$nm in height. For the RESET operation of the cell, an electrical current pulse of $50\,$ns is applied to the TEC with the BEC being grounded, so that current flows from top to bottom.

The proposed hybrid design structure is shown in Fig.~\ref{Fig1}(b), where a thin layer of SiGe acts as the bottom electrode material just below the GST and a vertical pillar PCM cell with a taper angle $\theta$ between the bottom electrode interface and the wall of the GST pillar. At $\theta=90^{\circ}$, the cell becomes the vertical pillar cell as given in Fig.~\ref{Fig1}(a), and at $\theta$=0$^{\circ}$, the cell assumes the mushroom structure. The taper angle is varied in our simulations to study the temperature profile and its subsequent effect on the programming current. 

\begin{figure}
	\centering
		\includegraphics[width=3.4in,height=2.8in]{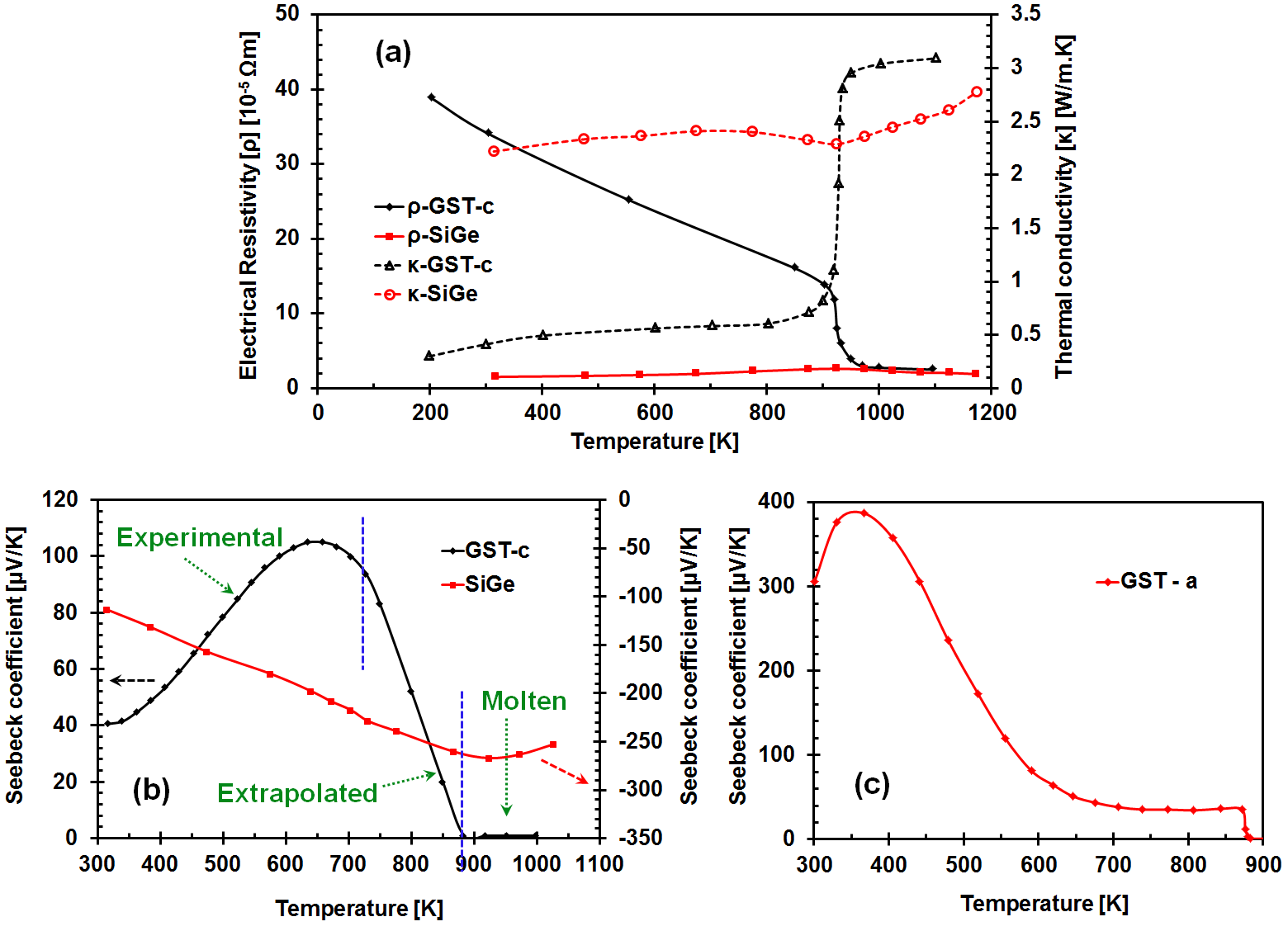}
		\caption{Temperature dependence of material properties. (a) Electrical resistivity and thermal conductivity variation for crystalline GST and SiGe. (b) Variation of the Seebeck coefficient of crystalline GST and SiGe with temperature. (c) Variation of the Seebeck coefficient of amorphous GST with temperature. For the molten state, a Seebeck coefficient of 0 $\mu$V/K is assumed.}
\label{Fig2}
\end{figure}

A finite element analysis of the structures is carried out using the COMSOL multi-physics package \cite{COMSOL} on a 2D axisymmetric geometry which solves the self consistent heat diffusion and current continuity equations given by 
\begin{equation}
{\rho'} C_p \frac{{\partial}T}{\partial t} = \nabla . (\kappa\nabla T) + \dot{Q}_{tot}
\end{equation}
\begin{equation}
\nabla . \textbf{J} = -\nabla . (\sigma (\nabla V + S \nabla T)),
\end{equation}
where $T$ and $V$ are the local temperature and electric potential respectively, ${\rho'}$ is the material density, $C_p$ is the heat capacity at constant pressure, $\kappa$ is thermal conductivity,  $\dot{Q}_{tot}$ is net rate of heat transfer rate per unit volume, $\textbf{J}$ is current density, $\sigma$ is the electrical conductivity, and $S$ is the Seebeck coefficient of the material. The above equation considers the effects due to the electrical and the heat current density \cite{Price}. 
\begin{equation}
\dot{Q}_{tot} = \dot{Q}_{J} + \dot{Q}_{T} + \dot{Q}_P
\end{equation}
\begin{equation}
\dot{Q}_{J} = \rho{\textbf{J}.\textbf{J}}
\end{equation}
\begin{equation}
\dot{Q}_{T} = -T{\frac {\partial S}{\partial T}}({\textbf{J}.\nabla T})
\label{Thomson}
\end{equation}
In the above equations, the thermoelectric Thomson effect acts as a source term $\dot{Q}_T$ in addition to the Joule heating term $\dot{Q}_{J}$. The additional Peltier term due to abrupt material interface at the GST-heater interface given by
\begin{equation}
\dot{Q}_{P} = -\textbf{J}.{\hat{z}}(T \Delta S)
\label{Peltier}
\end{equation}
which takes into account the abrupt difference in Seebeck coefficients across this interface. 

Temperature dependent material properties for GST and SiGe taken from corresponding literature are used in simulations \cite{bipin, S_GST, SiGe}, and are depicted in Fig.~\ref{Fig2}(b). The temperature dependence of the Seebeck coefficient of GST is plotted in Fig.~\ref{Fig2}(b). Here, data up to $740\,$K is taken from GST literature \cite{S_GST}, and from $740\,$K up to the melting point $873\,$K, the curve is extrapolated using standard thermoelectric models \cite{goldsmid} from solid state physics. The temperature variation of the Seebeck coefficient of amorphous GST is shown in Fig.~\ref{Fig2}(c). A value of $0\,\mu$V/K is assumed for the molten GST. Electrical resistivity and thermal conductivity of TiN is taken to be constant with temperature as $10^{-6}\,\Omega$m and 10 W/m.K respectively, and $1\,\Omega$m and 0.2 W/m.K for amorphous GST. Seebeck coefficient of TiN is taken to be 0 $\mu$V/K.

At the melting point, which is 873 K \cite{Yamada}, rapid quenching converts this molten region into amorphous. For the amorphous region, we hence take the entire molten region keeping a 5 K latent heat buffer to account for the molten-crystalline interface. This amorphous phase being highly resistive has a resistivity contrast of $10^{2}$ to $10^{4}$ from the conductive crystalline phase, and thus may be used in storing binary information. For reading, a short duration voltage pulse ($20\,$ns) small enough ($0.01\,$V) not to disturb the temperature of the cell is applied across the electrodes and the cell resistance (R$_{reset}$) is estimated via terminal current measurement. The cell is then said to be programmed and programming current is noted when the R$_{reset}$ to R$_{set}$ ratio reaches a value of $100$.
\section{Results}
Given that the hybrid structure proposed here has two prominent asymmetry aspects, we carry out a sequential exposition to understand the composite action of both aspects. 
\begin{figure}
	\centering
		\includegraphics[width=3.6in,height=2.8in]{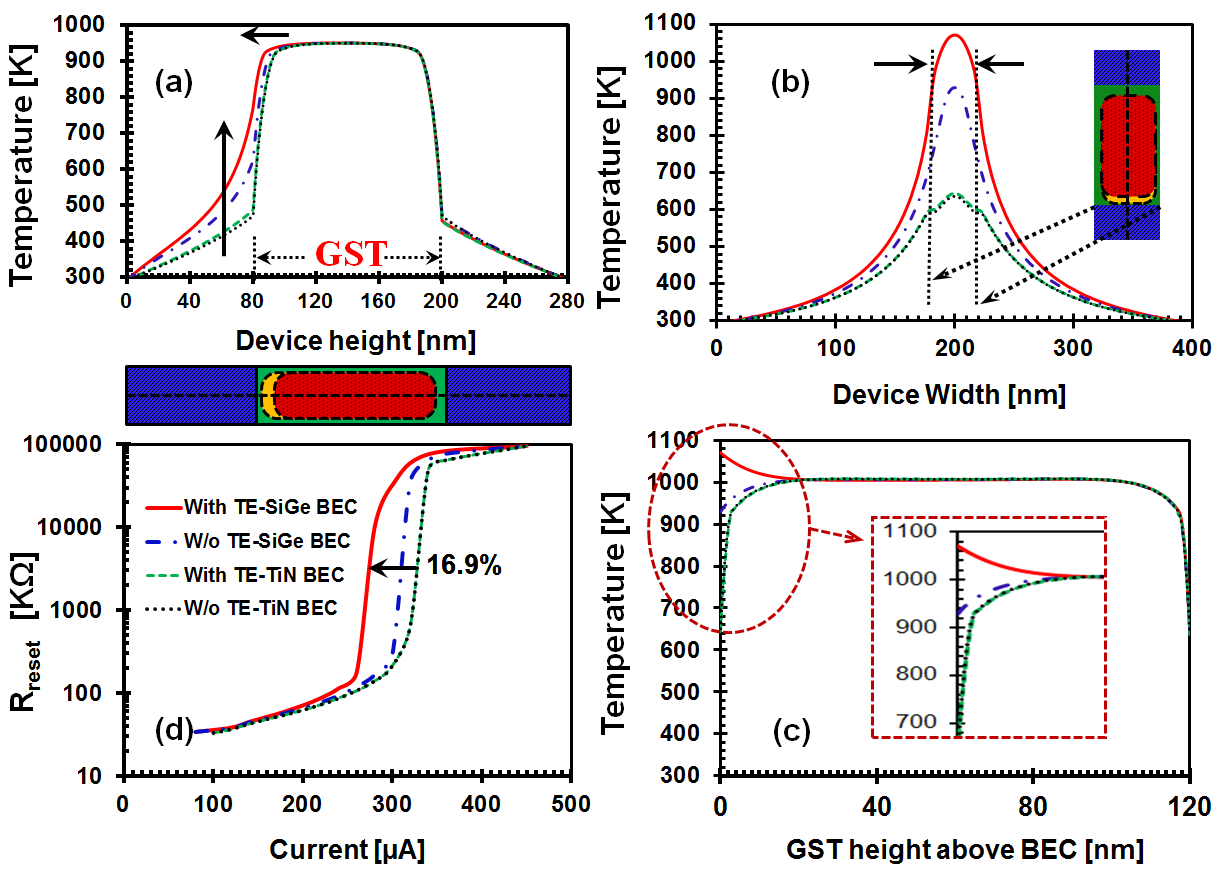}
		\caption{The effect of SiGe contact. Clockwise, (a) Central axis temperature profile of the vertical pillar PCM cell (shown in the corresponding schematic below) showing the shift of amorphous region (red and yellow) towards the bottom electrode due to enhanced thermoelectric effects at a current pulse amplitude of $140\,\mu$A. (b) Temperature profile at the GST-bottom electrode interface at $280\,\mu$A, the programming current for the SiGe case, shows an increase in interface temperature due to thermoelectric effects. (c) Corresponding temperature profile in the GST at the central axis at the programming current . (d) Effect of SiGe on programming current depicted by a variation of $R_{reset}$ with current magnitude. A common legend shown in (d) is to used for all sub plots.}
\label{Fig3}
\end{figure}

\subsection{Effect of SiGe Contact} 

The active region (GST) has a positive Seebeck coefficient implying that heat flows in the same direction as that of the charge carriers. It can be seen from \eqref{Peltier} that if a negative Seebeck coefficient material is used as the bottom electrode, it will increase the $\Delta S$ and hence the heat generated at the interface. A typical bottom electrode should have a high resistivity, low thermal conductivity and the ability to withstand high operating temperatures. Polycrystalline n-type SiGe is a well known thermoelectric material with a high negative Seebeck coefficient, with an operating range of $600^{\circ}$C to $1000^{\circ}$C \cite{SiGe}. Also, SiGe has much lower thermal conductivity than TiN which helps in minimizing the heat loss from the bottom electrode during the write operation \cite{SiGe_Lee_2006, SiGe_Lee_2008}. Hence n-type polycrystalline SiGe is our choice for the bottom electrode material.

The effect of using SiGe as a bottom electrode is shown in Fig.~\ref{Fig3}(a), where the temperature profile of the central axis at an input current of $140\,\mu$A is shown. This is much less than programming current, but is chosen to depict the onset of temperature variation at bottom electrode due to SiGe. Due to an additional Peltier heat generated, the molten hot spot shifts downwards, which, upon quenching, forms an amorphous region covering the interface completely. Hence, a very small current can flow through this amorphous region during the read operation. The black curve shows the temperature profile when TiN is used without the inclusion of thermoelectric effects resulting in a symmetric temperature profile. The inclusion of thermoelectric effects with TiN shows only a slight shift towards the bottom electrode (green dashed curve), which is not that impactful. But, with the use of SiGe as the bottom electrode, the amorphous region shifts downwards due to its higher resistivity and lower thermal conductivity than TiN, as seen in the blue dotted curve. With the inclusion of thermoelectric effects, the shift is greater (red solid curve), resulting in an overall decrease in programing current.

In Fig.~\ref{Fig3}(b), we show the temperature profile at the bottom electrode-GST interface with SiGe and TiN electrodes at the programming current, $280\,\mu$A, of the vertical cell. The interface has attained a temperature $T> 878$K at this current, whereas for TiN, the temperature attained is much smaller. At an increased current value, the temperature profile at the central axis also changes as shown in Fig.~\ref{Fig3}(c). Due to a greater influence of thermoelectric effects, a reduction of $16.9\%$ in the programming current is achieved, even for a vertical cell (Fig.~\ref{Fig3}(d)).
\begin{figure}
	\centering
		\includegraphics[width=3.5in,height=1.5in]{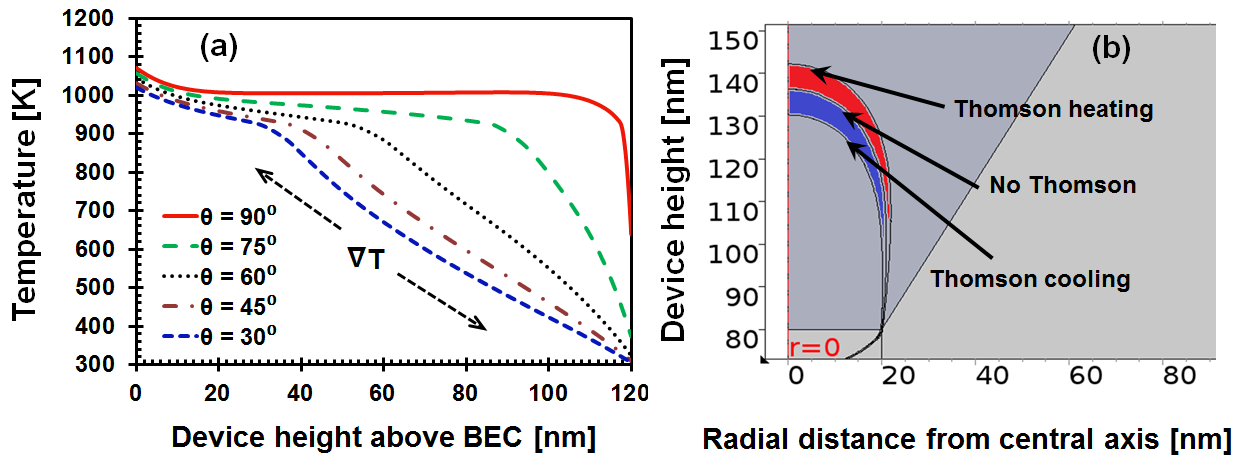}
		\caption{Effect of varying taper angle. (a) Temperature gradient in GST for various taper angles with SiGe as the bottom electrode, keeping the bottom contact radius as 20 nm and the height as 120 nm. The top electrode radius is varied to change the angle $\theta$. With an increase in taper, the temperature gradient $\nabla T$ increases resulting in increased Thomson effect. (b) Contour plot to show the change in amorphous volume as observed due to the Thomson effect contribution. Red region showing the increase in amorphous volume due to Thomson heating and blue region corresponds to Thomson cooling when a reverse polarity pulse is applied.}
\label{Fig4}
\end{figure}

\subsection{Effect of taper angle}
It is known that an increase in the aspect ratio decreases the programming current \cite{Liu} . But the fabrication of high aspect ratio vertical structures is not easy due to issues in the chemical vapor deposition process. Hence, a tapered vertical structure helping in high resolution pattering is easier to fabricate \cite{raoux, Confined}. Apart from this, the angle of the structure further helps to pull down the amorphous region towards the bottom electrode contact, as well as produces an increased Thomson effect, when SiGe is used as the bottom electrode. The volume of the amorphous region in the RESET state is also decreased in such a case. It is also well known that  in a PCM device, the resistance of the amorphous material increases steadily with power law dependence with time \cite{Reliability_2004, Reliability_2008}, which in turn affects the long-term cell reliability. Hence, a smaller volume of the amorphous region required for switching is always favorable to minimize this effective resistance drift. 

The effect of taper angle on the temperature profile of the central axis at $280\,\mu$A is shown in Fig.~\ref{Fig4}(a). With decrease in angle, the temperature gradient at the GST bulk increases, which when coupled with the temperature dependent Seebeck coefficient causes a net increase in the amorphous volume. At the operating range, which is around $873\,$K, the gradient, $dS/dT$ combined with the temperature gradient gives rise to an additional Thomson heat component (as noted in \eqref{Thomson}), and hence results in an increase in the volume (red region) along the direction of the temperature gradient ($\hat{z}$-direction), as noted in Fig.~\ref{Fig4}(b). The lower blue region depicts the effect of using a reverse polarity which results in Thomson cooling and hence a decrease in the amorphous volume.

\begin{figure}
	\centering
		\includegraphics[width=3.5in,height=1.4in]{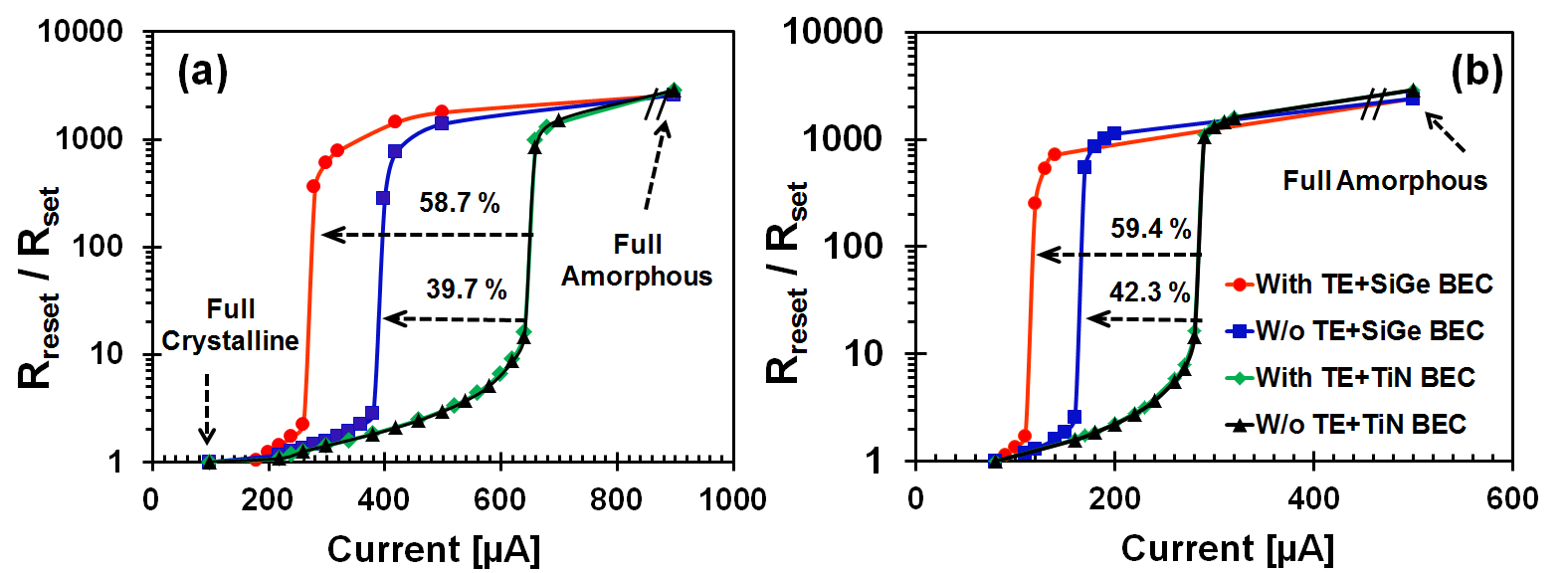}
		\caption{Structural impact on programming currents. R$_{reset}$ to R$_{set}$ ratio vs input current pulse amplitude for a $60^{\circ}$ structure with TiN and SiGe as bottom electrode, for (a) a $20\,$nm radius and (b) a $10\,$nm radius device. Programming current is reduced to $58.7 \%$ and $59.4 \%$ in the case of the $20\,$nm and the $10\,$nm devices respectively.}
\label{Fig5}
\end{figure}

\subsection{Role of Thermoelectric Effects} 
A comparison of the two 60$^{\circ}$ tapered structures, one with TiN as bottom electrode, and the other with SiGe as bottom electrode is carried out to clearly demonstrate the role of thermoelectric effects on the programming current. In Fig.~\ref{Fig5}(a), we plot $R_{reset}/R_{set}$ versus the input current for a device with radius of $20\,$nm and GST height of $120\,$nm, with $80\,$nm being the bottom electrode height. The programming current is obtained by interpolating between two points on the linear curve at which $R_{reset}/R_{set}=100$. Fig.~\ref{Fig5}(b) shows the similar plot for a device with a radius of $10\,$nm. With reduced radius, the programming current is reduced to half as compared to that of a 20nm radius cell. With TiN, thermoelectric effects play a minor role and hence there is a negligible reduction in the programming current. But with SiGe, a programming current reduction of around $40\%$ is seen due to the thermal conductivity reduction and a $60\%$ reduction is observed in conjunction with thermoelectric effects, in both the cases. Now that we have singled out the role of thermoelectric effects, for the upcoming sections of the paper, thermoelectric effects are included by default. 

\begin{figure}
	\centering
		\includegraphics[width=3.5in,height=1.6in]{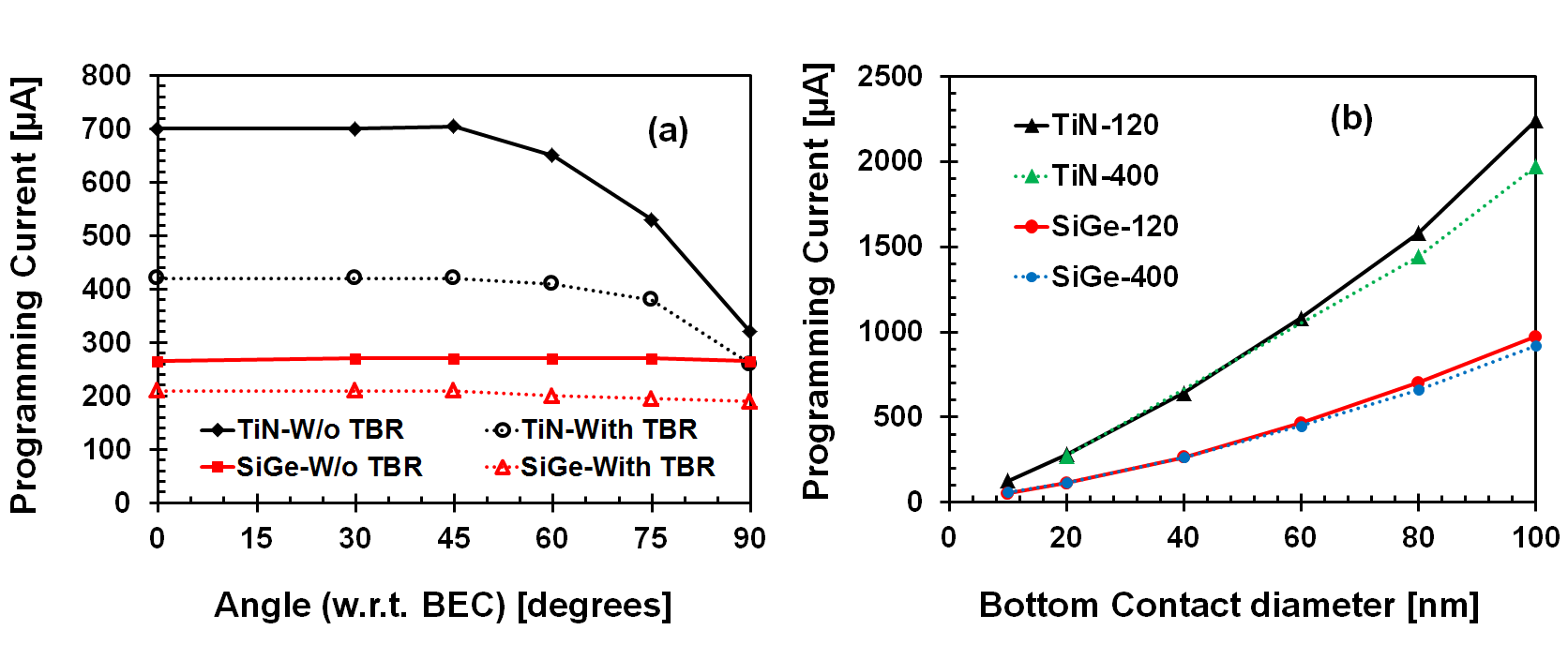}
		\caption{Effect of taper angle on programming currents. (a) Effect of using SiGe vs TiN as bottom electrode on programming currents as a function of varying angle for a cell with a $20\,$nm BEC radius and $120\,$nm GST height. With TiN electrode, the current varies from mushroom (0$^{\circ}$) to vertical (90$^{\circ}$) due to geometrical effects but with SiGe electrode, due to thermoelectric effects, the programming current is much lower and is almost independent of the contact angle of the pillar. Dotted lines show the effect of TBR along with Thermoelectric effects (b) Effect of increase in bottom electrode diameter on the programming current for a $60^{\circ}$ structure with SiGe as the bottom electrode for $120\,$nm and $400\,$nm GST heights.}
\label{Fig6}
\end{figure}

The effect of angle on programming currents is shown in Fig.~\ref{Fig6}(a), where we simulate a cell with a $20\,$nm BEC radius and $120\,$nm GST, with SiGe. In the case of the TiN structure, when the shape of amorphous region is hemispherical, the current remains constant with increase in angle from 0$^{\circ}$ to 45$^{\circ}$. However, with further increase in angle from 45$^{\circ}$ to 90$^{\circ}$, a sharp reduction in current is observed due to volume confinement. Hence, there is a substantial decrease in programming current as we move from 0$^{\circ}$ to 90$^{\circ}$, i.e., from a mushroom to a vertical structure. But in the case of SiGe BEC structure, thermoelectric effects alone are strong enough to override the current lowering effects of volume confinement. As the hot spot shifts towards the bottom electrode with increase in taper angle, more heat will dissipate through the bottom electrode. Hence, with further increase of angle, the current does not decrease for the SiGe case. Hence, the predicted lowering of programming currents is resilient to fabrication or process induced variability involved in creating the shaped cell, such as, uncontrolled tapering of the sidewalls due to etch induced modifications \cite{Etch}. So, effectively only the bottom electrode diameter i.e., contact diameter is left as the control parameter, thereby resulting in an ease of design in fabrication. \\ 
\indent It is important to note that our results up to now have not included the thermal boundary resistance (TBR) of the GST-BEC interface \cite{Reifenberg}. Here, the TBR and Peltier effect act in conjunction, whose physics merits a separate study \cite{Goodson}. However, a sample simulation with an assumed TBR value of 30 $m^{2}.K/GW$ for the GST-SiGe interface is shown in the dotted curve of \ref{Fig6}(a), demonstrating a further reduction in the programming current. 

We explore the effect of increase in bottom electrode radius in Fig.~\ref{Fig6}(b), for two different GST heights. With increase in the aspect ratio, i.e., increase in the GST height, the programming current reduces. The variation is much more in the case of TiN as the bottom electrode, but much less in the SiGe case.

\subsection{Effect of varying SiGe layer thickness}
\begin{figure}
	\centering
		\includegraphics[width=3.4in,height=2.8in]{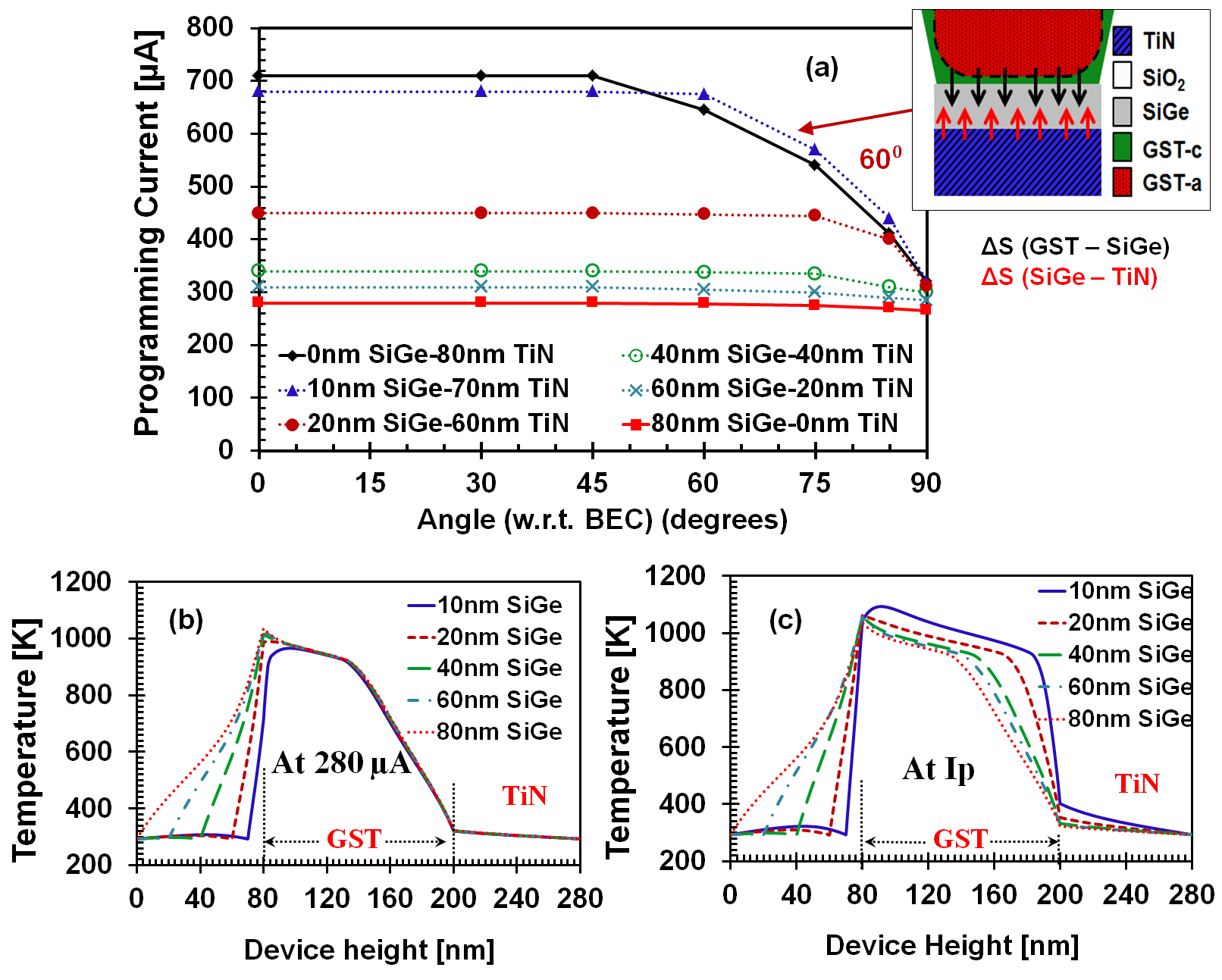}
		\caption{Role of varying the thickness of SiGe. (a) Effect of angle on programming current for various thicknesses of SiGe layer with GST height as 120nm and bottom electrode radius of 20nm. With smaller thicknesses of SiGe, counteracting thermoelectric effects due to the lower interface (shown in inset schematic) contribute to ineffectiveness in programming current reduction. (b) Central axis temperature profile of the cell  with different thickness of SiGe as bottom electrode (left region) keeping the total height of the electrode as 80nm. The temperature gradient is steeper with smaller thickness of SiGe, at $280\,\mu$A. (c) Temperature profile at their respective Programming currents.}
\label{Fig7}
\end{figure}

To understand the need for the bottom electrode to be completely made of SiGe, we now study the effects of varying the composition of the layered
bottom electrodes. The bottom electrode here consists of SiGe above TiN to
give a total electrode thickness of $80\,$nm. We vary the thickness of the SiGe layer from $10\,$nm to $80\,$nm and study the effect of angle on the
programming currents for the same GST height of $120\,$nm and bottom radius of $20\,$nm. The
results are shown in Fig.~\ref{Fig7}(a), which depicts the general trend with angle for different layer thicknesses. It is to be noted that for the $80\,$nm thick layer, the programming currents are relatively unaffected by the angle.

To explain this curious trend, one needs to delve into the physics of thermoelectric effects. We must note two crucial points, namely that as the angle decreases, the hot spot in the active region will move toward the GST-SiGe interface, and hence, the bottom electrode region could potentially become a heat sink at smaller angles, thus promoting an increase in the programming current as will be explained below.

We depict in Fig.~\ref{Fig7}(b), the temperature profiles for different SiGe layer thicknesses for a 60$^\circ$ angle with a $20\,$nm GST
radius and $120\,$nm height at $280\,\mu$A, which is programing current for full SiGe
electrode device. Fig.~\ref{Fig7}(c) shows the temperature
profiles for the same device at their respective programming currents. 

In the case of a very thin ($10\,$nm) SiGe layer, the Seebeck coefficient of the electrode
metal (TiN) is negligible while that of the n-type polycrystalline SiGe is highly negative. Therefore $\Delta S$ for the lower SiGe-TiN interface becomes negative as shown by the red arrows in the inset of Fig 7(a). This results in an opposite Peltier effect at the SiGe-TiN interface in the direction of current flow \eqref{Peltier}, which will in turn result in an enhanced cooling. This will make the region a good heat sink, thereby reducing the Peltier heating effect of the GST-SiGe layer due to the closer proximity with SiGe-TiN interface. Also, due to a sharp decrease in temperature from GST-SiGe interface to SiGe-TiN interface, the
temperature gradient in the narrow SiGe region will be
higher. In the case of a mushroom cell, as the hot spot is very close to the
interface, the temperature just below the GST-SiGe interface is very high ($> 900$K).
Given the temperature dependent Seebeck coefficient of SiGe \cite{SiGe}, $dS/dT$ of SiGe is positive and as dictated by \eqref{Thomson}, Thomson heating occurs which compensates for
the Peltier cooling caused by the lower interface to some extent. The net effect is an
additional heat generation and hence a smaller current.

With increase in angle from 45$^{\circ}$ to 90$^{\circ}$ (vertical), the hot
spot and hence the amorphous region will shift away
from GST-BEC interface. Hence the temperature near the
interface will be less than $900\,$K but the temperature gradient is still larger, which when
combined with the negative $dS/dT$ of SiGe \cite{SiGe}, results in Thomson cooling within the active region. Hence the hot spot is away from the SiGe-GST interface as shown by peak of the blue curve in Fig 7(b) and (c) for a $60^{\circ}$ taper. So due to both thermoelectric effects resulting in
cooling near the interface, a larger programming current than that using TiN-BEC structure
will be required, as shown in the blue dotted curve in the Fig.~\ref{Fig7}(a).

With increasing SiGe layer thickness, the Peltier cooling effect discussed above reduces due
to a greater separation between the SiGe-TiN and the GST-SiGe interfaces, as well as
decreased Thomson cooling caused by a decrease in the temperature gradient. But the
cooling effect is still not canceled completely and a higher current is required to
program the cell as compared to the cell with only SiGe. We must also note that  with further increase in SiGe thickness, the effect of cooling due to SiGe-TiN interface subsides enough
to match the full SiGe bottom electrode. The trend shows that there exists a
critical thickness beyond which the cooling effect of SiGe-TiN layer will be
negligible in comparison to GST-SiGe interface.
\begin{figure}
	\centering
		\includegraphics[width=3.5in,height=2.4in]{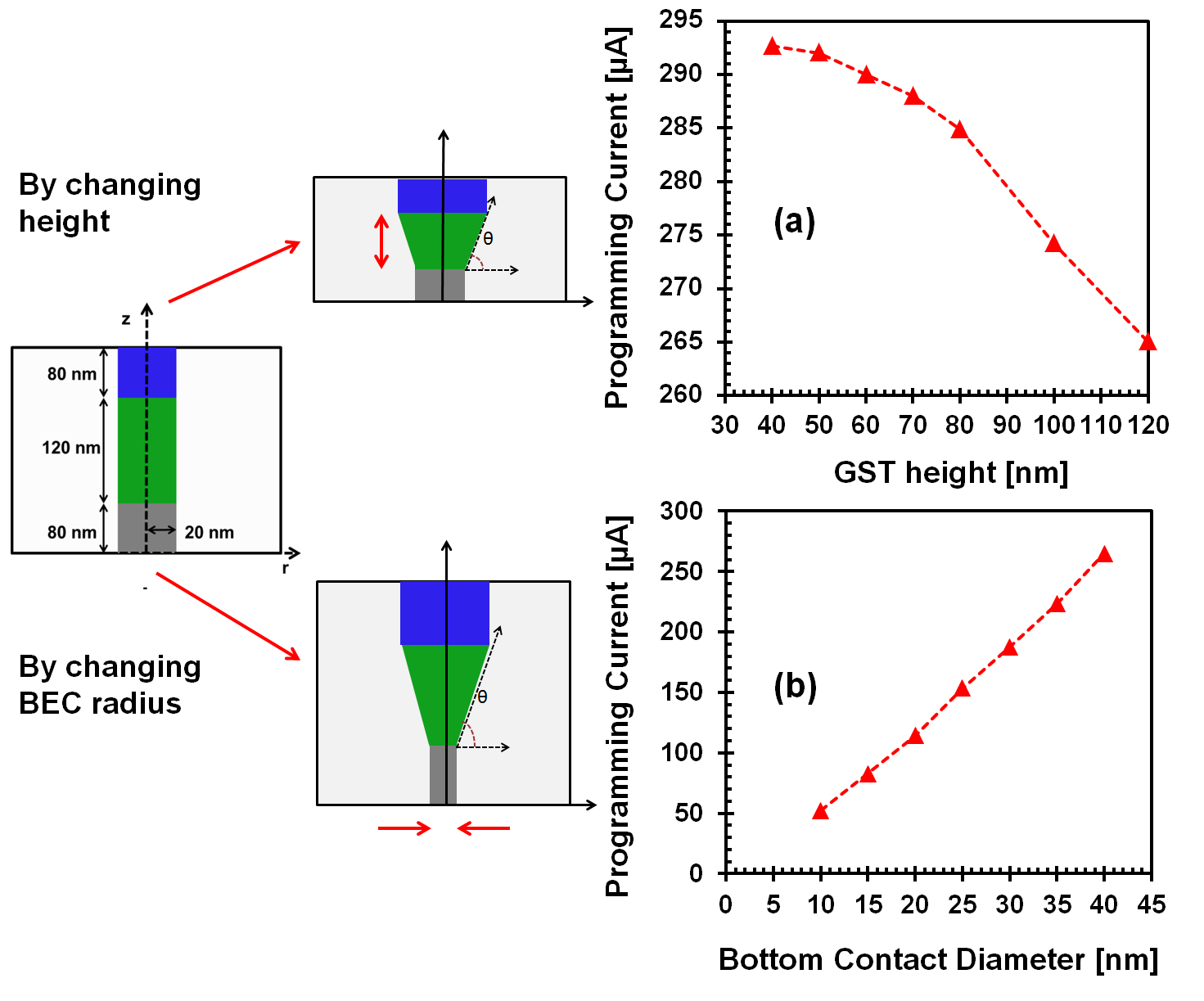}
		\caption{Isovolumetric studies. Effect of changing the height and bottom contact diameter on programming currents for an iso-volume cell. (a) With decrease in height, the programming current increases due to increased heat dissipation through the top electrode. (b) With a decrease in radius the current decreases linearly with bottom contact diameter.}
\label{Fig8}
\end{figure}
\subsection{Isovolumetric studies}
In order to understand the effect due to additional heat dissipation due to increased volume, we now carry out iso-volume studies. A tapered structure of the same volume as that of a pillar structure can be obtained either by reducing the GST height or by decreasing the bottom electrode radius. The results are shown in Fig.~\ref{Fig8}. With decrease in height of GST from $120\,$nm to $40\,$nm for a $20\,$nm bottom electrode radius, an increase of $\sim30\,\mu$A in the programming current is observed. With a shorter GST height, the top electrode is in the vicinity of the amorphous volume and heat starts dissipating through top electrode which was otherwise thermally adiabatic to the hot spot. But with decrease in the bottom electrode radius, the programming current reduces as also noted in the Fig.~\ref{Fig8}(b).

 
% \hfill September 17, 2014

%\IEEEpubidadjcol

\section{Conclusion}
We have proposed a new hybrid structural variant of the vertical pillar PCM cell which features a two-fold asymmetry to enhance the role played by thermoelectric effects on the programming current. Various device geometries were analyzed using 2D-axis-symmetric simulations to predict the effect on programming currents in the RESET operation, as well as for different thicknesses of the interface layer. A programming current reduction of up to $60\%$ was predicted depending upon the cell dimensions and geometry. It was also demonstrated that, remarkably, due to an interplay of Thomson cooling in the electrode and the asymmetric heating profile inside the active region, the predicted programming current reduction is resilient to fabrication variability. As the device dimensions shrink, it would be a fruitful venture to delve into a quantum transport treatment \cite{Lake, Akshay} of nanowire based PCM devices \cite{Liu}, so as to provide accurate models for futuristic PCM structures. 
\section*{Acknowledgment}
This work was supported in part by the IIT Bombay SEED grant and in part by the Center of Automotive Energy and Materials, ARCI, IIT Madras Research Park.

\bibliographystyle{apsrev}
\bibliography{ref_PCM}

\begin{thebibliography}{39}
\expandafter\ifx\csname natexlab\endcsname\relax\def\natexlab#1{#1}\fi
\expandafter\ifx\csname bibnamefont\endcsname\relax
  \def\bibnamefont#1{#1}\fi
\expandafter\ifx\csname bibfnamefont\endcsname\relax
  \def\bibfnamefont#1{#1}\fi
\expandafter\ifx\csname citenamefont\endcsname\relax
  \def\citenamefont#1{#1}\fi
\expandafter\ifx\csname url\endcsname\relax
  \def\url#1{\texttt{#1}}\fi
\expandafter\ifx\csname urlprefix\endcsname\relax\def\urlprefix{URL }\fi
\providecommand{\bibinfo}[2]{#2}
\providecommand{\eprint}[2][]{\url{#2}}

\bibitem[{\citenamefont{Raoux and Wuttig}(2009)}]{raoux}
\bibinfo{author}{\bibfnamefont{S.}~\bibnamefont{Raoux}} \bibnamefont{and}
  \bibinfo{author}{\bibfnamefont{M.}~\bibnamefont{Wuttig}},
  \emph{\bibinfo{title}{Phase {C}hange {M}aterials: {S}cience and
  {A}pplications}} (\bibinfo{publisher}{Springer}, \bibinfo{year}{2009}).

\bibitem[{\citenamefont{Qureshi et~al.}(2012)\citenamefont{Qureshi, Gurumurthi,
  and Rajendran}}]{bipin}
\bibinfo{author}{\bibfnamefont{M.~K.} \bibnamefont{Qureshi}},
  \bibinfo{author}{\bibfnamefont{S.}~\bibnamefont{Gurumurthi}},
  \bibnamefont{and}
  \bibinfo{author}{\bibfnamefont{B.}~\bibnamefont{Rajendran}},
  \emph{\bibinfo{title}{Phase {C}hange {M}emory: {F}rom {D}evices to
  {S}ystems}} (\bibinfo{publisher}{Morgan and Claypool}, \bibinfo{year}{2012}).

\bibitem[{\citenamefont{Burr et~al.}(2010)\citenamefont{Burr, Breitwisch,
  Franceschini, Garetto, Gopalakrishnan, Jackson, Kurdi, Lam, Lastras, Padilla
  et~al.}}]{Burr}
\bibinfo{author}{\bibfnamefont{G.~W.} \bibnamefont{Burr}},
  \bibinfo{author}{\bibfnamefont{M.~J.} \bibnamefont{Breitwisch}},
  \bibinfo{author}{\bibfnamefont{M.}~\bibnamefont{Franceschini}},
  \bibinfo{author}{\bibfnamefont{D.}~\bibnamefont{Garetto}},
  \bibinfo{author}{\bibfnamefont{K.}~\bibnamefont{Gopalakrishnan}},
  \bibinfo{author}{\bibfnamefont{B.}~\bibnamefont{Jackson}},
  \bibinfo{author}{\bibfnamefont{B.}~\bibnamefont{Kurdi}},
  \bibinfo{author}{\bibfnamefont{C.}~\bibnamefont{Lam}},
  \bibinfo{author}{\bibfnamefont{L.~A.} \bibnamefont{Lastras}},
  \bibinfo{author}{\bibfnamefont{A.}~\bibnamefont{Padilla}},
  \bibnamefont{et~al.}, \bibinfo{journal}{J. Vac. Sci. Technol. B}
  \textbf{\bibinfo{volume}{28}}, \bibinfo{pages}{223} (\bibinfo{year}{2010}).

\bibitem[{\citenamefont{Wong et~al.}(2010)\citenamefont{Wong, Raoux, Kim,
  Liang, Reifenberg, Rajendran, Asheghi, and Goodson}}]{Wong}
\bibinfo{author}{\bibfnamefont{H.~P.} \bibnamefont{Wong}},
  \bibinfo{author}{\bibfnamefont{S.}~\bibnamefont{Raoux}},
  \bibinfo{author}{\bibfnamefont{S.}~\bibnamefont{Kim}},
  \bibinfo{author}{\bibfnamefont{J.}~\bibnamefont{Liang}},
  \bibinfo{author}{\bibfnamefont{J.~P.} \bibnamefont{Reifenberg}},
  \bibinfo{author}{\bibfnamefont{B.}~\bibnamefont{Rajendran}},
  \bibinfo{author}{\bibfnamefont{M.}~\bibnamefont{Asheghi}}, \bibnamefont{and}
  \bibinfo{author}{\bibfnamefont{K.~E.} \bibnamefont{Goodson}},
  \bibinfo{journal}{Proc. IEEE} \textbf{\bibinfo{volume}{98}},
  \bibinfo{pages}{2201} (\bibinfo{year}{2010}).

\bibitem[{\citenamefont{Happ et~al.}(2006)\citenamefont{Happ, Breitwisch,
  Schrott, Philipp, Lee, Cheek, Nirschl, Lamorey, Ho, Chen et~al.}}]{Pillar}
\bibinfo{author}{\bibfnamefont{T.~D.} \bibnamefont{Happ}},
  \bibinfo{author}{\bibfnamefont{M.}~\bibnamefont{Breitwisch}},
  \bibinfo{author}{\bibfnamefont{A.}~\bibnamefont{Schrott}},
  \bibinfo{author}{\bibfnamefont{J.~B.} \bibnamefont{Philipp}},
  \bibinfo{author}{\bibfnamefont{M.~H.} \bibnamefont{Lee}},
  \bibinfo{author}{\bibfnamefont{R.}~\bibnamefont{Cheek}},
  \bibinfo{author}{\bibfnamefont{T.}~\bibnamefont{Nirschl}},
  \bibinfo{author}{\bibfnamefont{M.}~\bibnamefont{Lamorey}},
  \bibinfo{author}{\bibfnamefont{C.~H.} \bibnamefont{Ho}},
  \bibinfo{author}{\bibfnamefont{S.~H.} \bibnamefont{Chen}},
  \bibnamefont{et~al.}, \bibinfo{journal}{Proc. Symp. VLSI Technol.} pp.
  \bibinfo{pages}{120--121} (\bibinfo{year}{2006}).

\bibitem[{\citenamefont{Breitwisch et~al.}(2007)\citenamefont{Breitwisch,
  Nirschl, Chen, Zhu, Lee, Lamorey, Burr, Joseph, Schrott, Philipp
  et~al.}}]{Pore}
\bibinfo{author}{\bibfnamefont{M.}~\bibnamefont{Breitwisch}},
  \bibinfo{author}{\bibfnamefont{T.}~\bibnamefont{Nirschl}},
  \bibinfo{author}{\bibfnamefont{C.~F.} \bibnamefont{Chen}},
  \bibinfo{author}{\bibfnamefont{Y.}~\bibnamefont{Zhu}},
  \bibinfo{author}{\bibfnamefont{M.~H.} \bibnamefont{Lee}},
  \bibinfo{author}{\bibfnamefont{M.}~\bibnamefont{Lamorey}},
  \bibinfo{author}{\bibfnamefont{G.~W.} \bibnamefont{Burr}},
  \bibinfo{author}{\bibfnamefont{E.}~\bibnamefont{Joseph}},
  \bibinfo{author}{\bibfnamefont{A.}~\bibnamefont{Schrott}},
  \bibinfo{author}{\bibfnamefont{J.~B.} \bibnamefont{Philipp}},
  \bibnamefont{et~al.}, \bibinfo{journal}{Proc. Symp. VLSI Technol.} pp.
  \bibinfo{pages}{100--101} (\bibinfo{year}{2007}).

\bibitem[{\citenamefont{Cho et~al.}(2005)\citenamefont{Cho, Yi, Ha, Kuh, Lee,
  Park, Nam, Horii, Cho, Ryoo et~al.}}]{Cho}
\bibinfo{author}{\bibfnamefont{S.~L.} \bibnamefont{Cho}},
  \bibinfo{author}{\bibfnamefont{J.~H.} \bibnamefont{Yi}},
  \bibinfo{author}{\bibfnamefont{Y.~H.} \bibnamefont{Ha}},
  \bibinfo{author}{\bibfnamefont{B.~J.} \bibnamefont{Kuh}},
  \bibinfo{author}{\bibfnamefont{C.~M.} \bibnamefont{Lee}},
  \bibinfo{author}{\bibfnamefont{J.~H.} \bibnamefont{Park}},
  \bibinfo{author}{\bibfnamefont{S.~D.} \bibnamefont{Nam}},
  \bibinfo{author}{\bibfnamefont{H.}~\bibnamefont{Horii}},
  \bibinfo{author}{\bibfnamefont{B.~O.} \bibnamefont{Cho}},
  \bibinfo{author}{\bibfnamefont{K.~C.} \bibnamefont{Ryoo}},
  \bibnamefont{et~al.}, \bibinfo{journal}{Symp. VLSI Tech.} pp.
  \bibinfo{pages}{96--97} (\bibinfo{year}{2005}).

\bibitem[{\citenamefont{Bedeschi et~al.}(2005)\citenamefont{Bedeschi, Bez,
  Boffino, Bonizzoni, Buda, Casagrande, Costa, Ferraro, Gastaldi, Khouri
  et~al.}}]{Mu_Trench}
\bibinfo{author}{\bibfnamefont{F.}~\bibnamefont{Bedeschi}},
  \bibinfo{author}{\bibfnamefont{R.}~\bibnamefont{Bez}},
  \bibinfo{author}{\bibfnamefont{C.}~\bibnamefont{Boffino}},
  \bibinfo{author}{\bibfnamefont{E.}~\bibnamefont{Bonizzoni}},
  \bibinfo{author}{\bibfnamefont{E.~C.} \bibnamefont{Buda}},
  \bibinfo{author}{\bibfnamefont{G.}~\bibnamefont{Casagrande}},
  \bibinfo{author}{\bibfnamefont{L.}~\bibnamefont{Costa}},
  \bibinfo{author}{\bibfnamefont{M.}~\bibnamefont{Ferraro}},
  \bibinfo{author}{\bibfnamefont{R.}~\bibnamefont{Gastaldi}},
  \bibinfo{author}{\bibfnamefont{O.}~\bibnamefont{Khouri}},
  \bibnamefont{et~al.}, \bibinfo{journal}{IEEE Journ of Solid State Circ.}
  \textbf{\bibinfo{volume}{40}}, \bibinfo{pages}{1557} (\bibinfo{year}{2005}).

\bibitem[{\citenamefont{Lankhorst et~al.}(2005)\citenamefont{Lankhorst,
  Ketelaars, and Wolters}}]{Line}
\bibinfo{author}{\bibfnamefont{M.~H.~R.} \bibnamefont{Lankhorst}},
  \bibinfo{author}{\bibfnamefont{B.~W. S. M.~M.} \bibnamefont{Ketelaars}},
  \bibnamefont{and} \bibinfo{author}{\bibfnamefont{R.~A.~M.}
  \bibnamefont{Wolters}}, \bibinfo{journal}{Nature Mater.}
  \textbf{\bibinfo{volume}{4}}, \bibinfo{pages}{347} (\bibinfo{year}{2005}).

\bibitem[{\citenamefont{Chen et~al.}(2006)\citenamefont{Chen, Rettner, Raoux,
  Burr, Chen, Shelby, Salinga, Risk, Happ, McCleland et~al.}}]{Bridge}
\bibinfo{author}{\bibfnamefont{Y.~C.} \bibnamefont{Chen}},
  \bibinfo{author}{\bibfnamefont{C.~T.} \bibnamefont{Rettner}},
  \bibinfo{author}{\bibfnamefont{S.}~\bibnamefont{Raoux}},
  \bibinfo{author}{\bibfnamefont{G.~W.} \bibnamefont{Burr}},
  \bibinfo{author}{\bibfnamefont{S.~H.} \bibnamefont{Chen}},
  \bibinfo{author}{\bibfnamefont{R.~M.} \bibnamefont{Shelby}},
  \bibinfo{author}{\bibfnamefont{M.}~\bibnamefont{Salinga}},
  \bibinfo{author}{\bibfnamefont{W.~P.} \bibnamefont{Risk}},
  \bibinfo{author}{\bibfnamefont{T.~D.} \bibnamefont{Happ}},
  \bibinfo{author}{\bibfnamefont{G.~M.} \bibnamefont{McCleland}},
  \bibnamefont{et~al.}, \bibinfo{journal}{IEDM Tech. Dig.} pp.
  \bibinfo{pages}{777--780} (\bibinfo{year}{2006}).

\bibitem[{\citenamefont{Russo et~al.}(2008)\citenamefont{Russo, Ielmini,
  Redaelli, and Lacaita}}]{Russo}
\bibinfo{author}{\bibfnamefont{U.}~\bibnamefont{Russo}},
  \bibinfo{author}{\bibfnamefont{D.}~\bibnamefont{Ielmini}},
  \bibinfo{author}{\bibfnamefont{A.}~\bibnamefont{Redaelli}}, \bibnamefont{and}
  \bibinfo{author}{\bibfnamefont{A.~L.} \bibnamefont{Lacaita}},
  \bibinfo{journal}{IEEE Trans. Electr. Dev.} \textbf{\bibinfo{volume}{55}},
  \bibinfo{pages}{506} (\bibinfo{year}{2008}).

\bibitem[{\citenamefont{Chen and Pop}(2009)}]{Pop}
\bibinfo{author}{\bibfnamefont{I.-R.} \bibnamefont{Chen}} \bibnamefont{and}
  \bibinfo{author}{\bibfnamefont{E.}~\bibnamefont{Pop}},
  \bibinfo{journal}{Electron Devices, IEEE Transactions on}
  \textbf{\bibinfo{volume}{56}}, \bibinfo{pages}{1523} (\bibinfo{year}{2009}).

\bibitem[{\citenamefont{Liu et~al.}(2011)\citenamefont{Liu, Yu, and
  Anantram}}]{Liu}
\bibinfo{author}{\bibfnamefont{J.}~\bibnamefont{Liu}},
  \bibinfo{author}{\bibfnamefont{B.}~\bibnamefont{Yu}}, \bibnamefont{and}
  \bibinfo{author}{\bibfnamefont{M.~P.} \bibnamefont{Anantram}},
  \bibinfo{journal}{IEEE Elec. Dev. Letters} \textbf{\bibinfo{volume}{32}},
  \bibinfo{pages}{1340} (\bibinfo{year}{2011}).

\bibitem[{\citenamefont{Liu}(25~Mar.~2014)}]{BEC_via}
\bibinfo{author}{\bibfnamefont{J.}~\bibnamefont{Liu}},
  \bibinfo{howpublished}{{U.S. Patent No.} 8,679,934 B2}
  (\bibinfo{year}{25~Mar.~2014}).

\bibitem[{\citenamefont{Lai et~al.}(2005)\citenamefont{Lai, Qiao, Feng, Le, La,
  Lin, Tang, Cai, and Chen}}]{Nitrogen}
\bibinfo{author}{\bibfnamefont{Y.~F.} \bibnamefont{Lai}},
  \bibinfo{author}{\bibfnamefont{B.~W.} \bibnamefont{Qiao}},
  \bibinfo{author}{\bibfnamefont{J.}~\bibnamefont{Feng}},
  \bibinfo{author}{\bibfnamefont{Y.}~\bibnamefont{Le}},
  \bibinfo{author}{\bibfnamefont{L.~Z.} \bibnamefont{La}},
  \bibinfo{author}{\bibfnamefont{Y.~Y.} \bibnamefont{Lin}},
  \bibinfo{author}{\bibfnamefont{T.~A.} \bibnamefont{Tang}},
  \bibinfo{author}{\bibfnamefont{B.~C.} \bibnamefont{Cai}}, \bibnamefont{and}
  \bibinfo{author}{\bibfnamefont{B.~M.} \bibnamefont{Chen}},
  \bibinfo{journal}{J. Electron. Mater.} \textbf{\bibinfo{volume}{34}},
  \bibinfo{pages}{176} (\bibinfo{year}{2005}).

\bibitem[{\citenamefont{Lee et~al.}(2012)\citenamefont{Lee, Asheghi, and
  Goodson}}]{Goodson}
\bibinfo{author}{\bibfnamefont{J.}~\bibnamefont{Lee}},
  \bibinfo{author}{\bibfnamefont{M.}~\bibnamefont{Asheghi}}, \bibnamefont{and}
  \bibinfo{author}{\bibfnamefont{K.~E.} \bibnamefont{Goodson}},
  \bibinfo{journal}{Nanotechnology} \textbf{\bibinfo{volume}{23}},
  \bibinfo{pages}{205201} (\bibinfo{year}{2012}).

\bibitem[{\citenamefont{Faraclas et~al.}(2014)\citenamefont{Faraclas, Bakan,
  Adnane, Dirisaglik, Williams, Gokirmak, and Silva}}]{Azer}
\bibinfo{author}{\bibfnamefont{A.}~\bibnamefont{Faraclas}},
  \bibinfo{author}{\bibfnamefont{G.}~\bibnamefont{Bakan}},
  \bibinfo{author}{\bibfnamefont{L.}~\bibnamefont{Adnane}},
  \bibinfo{author}{\bibfnamefont{F.}~\bibnamefont{Dirisaglik}},
  \bibinfo{author}{\bibfnamefont{N.~E.} \bibnamefont{Williams}},
  \bibinfo{author}{\bibfnamefont{A.}~\bibnamefont{Gokirmak}}, \bibnamefont{and}
  \bibinfo{author}{\bibfnamefont{H.}~\bibnamefont{Silva}},
  \bibinfo{journal}{IEEE Trans. Electron Devices}
  \textbf{\bibinfo{volume}{61}}, \bibinfo{pages}{372} (\bibinfo{year}{2014}).

\bibitem[{\citenamefont{Suh et~al.}(2010)\citenamefont{Suh, Kim, Kim, Kang,
  Lee, Khang, Park, Yoon, Im, and Ihm}}]{Suh}
\bibinfo{author}{\bibfnamefont{D.-S.} \bibnamefont{Suh}},
  \bibinfo{author}{\bibfnamefont{C.}~\bibnamefont{Kim}},
  \bibinfo{author}{\bibfnamefont{K.~H.~P.} \bibnamefont{Kim}},
  \bibinfo{author}{\bibfnamefont{Y.-S.} \bibnamefont{Kang}},
  \bibinfo{author}{\bibfnamefont{T.-Y.} \bibnamefont{Lee}},
  \bibinfo{author}{\bibfnamefont{Y.}~\bibnamefont{Khang}},
  \bibinfo{author}{\bibfnamefont{T.~S.} \bibnamefont{Park}},
  \bibinfo{author}{\bibfnamefont{Y.-G.} \bibnamefont{Yoon}},
  \bibinfo{author}{\bibfnamefont{J.}~\bibnamefont{Im}}, \bibnamefont{and}
  \bibinfo{author}{\bibfnamefont{J.}~\bibnamefont{Ihm}},
  \bibinfo{journal}{Appl. Phys. Lett.} \textbf{\bibinfo{volume}{96}},
  \bibinfo{pages}{123115} (\bibinfo{year}{2010}).

\bibitem[{\citenamefont{Castro et~al.}(2007)\citenamefont{Castro, Goux, Hurkx,
  Attenborough, Delhougne, Lisoni, Jedema, Zandt, Wolters, and
  Gravesteijn}}]{Castro}
\bibinfo{author}{\bibfnamefont{D.~T.} \bibnamefont{Castro}},
  \bibinfo{author}{\bibfnamefont{L.}~\bibnamefont{Goux}},
  \bibinfo{author}{\bibfnamefont{G.~A.~M.} \bibnamefont{Hurkx}},
  \bibinfo{author}{\bibfnamefont{K.}~\bibnamefont{Attenborough}},
  \bibinfo{author}{\bibfnamefont{R.}~\bibnamefont{Delhougne}},
  \bibinfo{author}{\bibfnamefont{J.}~\bibnamefont{Lisoni}},
  \bibinfo{author}{\bibfnamefont{F.~J.} \bibnamefont{Jedema}},
  \bibinfo{author}{\bibfnamefont{M.~A.~A.} \bibnamefont{Zandt}},
  \bibinfo{author}{\bibfnamefont{R.~A.~M.} \bibnamefont{Wolters}},
  \bibnamefont{and} \bibinfo{author}{\bibfnamefont{D.~J.}
  \bibnamefont{Gravesteijn}}, \bibinfo{journal}{Proc. IEEE IEDM} pp.
  \bibinfo{pages}{315--318} (\bibinfo{year}{2007}).

\bibitem[{\citenamefont{Nirschl et~al.}(2007)\citenamefont{Nirschl, Philipp,
  Happ, Burr, Rajendran, Lee, Schrott, Yang, Breitwisch, Chen
  et~al.}}]{Nirschl}
\bibinfo{author}{\bibfnamefont{T.}~\bibnamefont{Nirschl}},
  \bibinfo{author}{\bibfnamefont{J.~B.} \bibnamefont{Philipp}},
  \bibinfo{author}{\bibfnamefont{T.~D.} \bibnamefont{Happ}},
  \bibinfo{author}{\bibfnamefont{G.~W.} \bibnamefont{Burr}},
  \bibinfo{author}{\bibfnamefont{B.}~\bibnamefont{Rajendran}},
  \bibinfo{author}{\bibfnamefont{M.-H.} \bibnamefont{Lee}},
  \bibinfo{author}{\bibfnamefont{A.}~\bibnamefont{Schrott}},
  \bibinfo{author}{\bibfnamefont{M.}~\bibnamefont{Yang}},
  \bibinfo{author}{\bibfnamefont{M.}~\bibnamefont{Breitwisch}},
  \bibinfo{author}{\bibfnamefont{C.-F.} \bibnamefont{Chen}},
  \bibnamefont{et~al.}, \bibinfo{journal}{IEDM Tech. Dig.} pp.
  \bibinfo{pages}{461--464} (\bibinfo{year}{2007}).

\bibitem[{\citenamefont{Kim and Hwang}(2009)}]{Kim}
\bibinfo{author}{\bibfnamefont{K.~M.} \bibnamefont{Kim}} \bibnamefont{and}
  \bibinfo{author}{\bibfnamefont{C.~S.} \bibnamefont{Hwang}},
  \bibinfo{journal}{Appl. Phys. Lett.} \textbf{\bibinfo{volume}{94}},
  \bibinfo{pages}{122109} (\bibinfo{year}{2009}).

\bibitem[{\citenamefont{Choi et~al.}(2010)\citenamefont{Choi, Choi, Eom, Rha,
  Kim, and Cheol}}]{Choi}
\bibinfo{author}{\bibfnamefont{B.~J.} \bibnamefont{Choi}},
  \bibinfo{author}{\bibfnamefont{S.}~\bibnamefont{Choi}},
  \bibinfo{author}{\bibfnamefont{T.}~\bibnamefont{Eom}},
  \bibinfo{author}{\bibfnamefont{S.~H.} \bibnamefont{Rha}},
  \bibinfo{author}{\bibfnamefont{K.~M.} \bibnamefont{Kim}}, \bibnamefont{and}
  \bibinfo{author}{\bibfnamefont{S.~H.} \bibnamefont{Cheol}},
  \bibinfo{journal}{Appl. Phys. Lett.} \textbf{\bibinfo{volume}{97}},
  \bibinfo{pages}{132107} (\bibinfo{year}{2010}).

\bibitem[{\citenamefont{Hubert et~al.}(2011)\citenamefont{Hubert, Jahan,
  Toffol, Perniola, Sousa, Persico, Nodin, Grampeix, Aussenac, and
  De~Salvo}}]{Hubert}
\bibinfo{author}{\bibfnamefont{Q.}~\bibnamefont{Hubert}},
  \bibinfo{author}{\bibfnamefont{C.}~\bibnamefont{Jahan}},
  \bibinfo{author}{\bibfnamefont{A.}~\bibnamefont{Toffol}},
  \bibinfo{author}{\bibfnamefont{L.}~\bibnamefont{Perniola}},
  \bibinfo{author}{\bibfnamefont{V.}~\bibnamefont{Sousa}},
  \bibinfo{author}{\bibfnamefont{A.}~\bibnamefont{Persico}},
  \bibinfo{author}{\bibfnamefont{J.}~\bibnamefont{Nodin}},
  \bibinfo{author}{\bibfnamefont{H.}~\bibnamefont{Grampeix}},
  \bibinfo{author}{\bibfnamefont{F.}~\bibnamefont{Aussenac}}, \bibnamefont{and}
  \bibinfo{author}{\bibfnamefont{B.}~\bibnamefont{De~Salvo}},
  \bibinfo{journal}{Proc. Solid-State Device Res. Conf. (ESSDERC), Eur.}
  \textbf{\bibinfo{volume}{61}}, \bibinfo{pages}{95} (\bibinfo{year}{2011}).

\bibitem[{\citenamefont{Wu et~al.}(2011)\citenamefont{Wu, Breitwisch, Kim, Hsu,
  Cheek, Du, Li, Lai, Zhu, Wang et~al.}}]{Wu}
\bibinfo{author}{\bibfnamefont{J.~Y.} \bibnamefont{Wu}},
  \bibinfo{author}{\bibfnamefont{M.}~\bibnamefont{Breitwisch}},
  \bibinfo{author}{\bibfnamefont{S.}~\bibnamefont{Kim}},
  \bibinfo{author}{\bibfnamefont{T.~H.} \bibnamefont{Hsu}},
  \bibinfo{author}{\bibfnamefont{R.}~\bibnamefont{Cheek}},
  \bibinfo{author}{\bibfnamefont{P.~Y.} \bibnamefont{Du}},
  \bibinfo{author}{\bibfnamefont{J.}~\bibnamefont{Li}},
  \bibinfo{author}{\bibfnamefont{E.~K.} \bibnamefont{Lai}},
  \bibinfo{author}{\bibfnamefont{Y.}~\bibnamefont{Zhu}},
  \bibinfo{author}{\bibfnamefont{T.~Y.} \bibnamefont{Wang}},
  \bibnamefont{et~al.}, \bibinfo{journal}{Proc. IEEE IEDM} pp.
  \bibinfo{pages}{3.2.1--3.2.4} (\bibinfo{year}{2011}).

\bibitem[{\citenamefont{Lee et~al.}(2006)\citenamefont{Lee, Choi, Ryu, Yoon,
  Lee, Park, Kim, Lee, and Yu}}]{SiGe_Lee_2006}
\bibinfo{author}{\bibfnamefont{S.-Y.} \bibnamefont{Lee}},
  \bibinfo{author}{\bibfnamefont{K.-J.} \bibnamefont{Choi}},
  \bibinfo{author}{\bibfnamefont{S.-O.} \bibnamefont{Ryu}},
  \bibinfo{author}{\bibfnamefont{S.-M.} \bibnamefont{Yoon}},
  \bibinfo{author}{\bibfnamefont{N.-Y.} \bibnamefont{Lee}},
  \bibinfo{author}{\bibfnamefont{Y.-S.} \bibnamefont{Park}},
  \bibinfo{author}{\bibfnamefont{S.-H.} \bibnamefont{Kim}},
  \bibinfo{author}{\bibfnamefont{S.-H.} \bibnamefont{Lee}}, \bibnamefont{and}
  \bibinfo{author}{\bibfnamefont{B.-G.} \bibnamefont{Yu}},
  \bibinfo{journal}{Appl. Phys. Lett.} \textbf{\bibinfo{volume}{89}},
  \bibinfo{pages}{053517} (\bibinfo{year}{2006}).

\bibitem[{\citenamefont{Lee et~al.}(2008)\citenamefont{Lee, Park, Yoon, Jung,
  and Yu}}]{SiGe_Lee_2008}
\bibinfo{author}{\bibfnamefont{S.-Y.} \bibnamefont{Lee}},
  \bibinfo{author}{\bibfnamefont{Y.-S.} \bibnamefont{Park}},
  \bibinfo{author}{\bibfnamefont{S.-M.} \bibnamefont{Yoon}},
  \bibinfo{author}{\bibfnamefont{S.-W.} \bibnamefont{Jung}}, \bibnamefont{and}
  \bibinfo{author}{\bibfnamefont{B.-G.} \bibnamefont{Yu}}, \bibinfo{journal}{J.
  Electrochem. Soc.} \textbf{\bibinfo{volume}{155}}, \bibinfo{pages}{H314}
  (\bibinfo{year}{2008}).

\bibitem[{\citenamefont{Lee et~al.}(2007)\citenamefont{Lee, Park, Cho, Park,
  Bae, Park, Park, An, Bae, Ahn et~al.}}]{Confined}
\bibinfo{author}{\bibfnamefont{J.~I.} \bibnamefont{Lee}},
  \bibinfo{author}{\bibfnamefont{H.}~\bibnamefont{Park}},
  \bibinfo{author}{\bibfnamefont{S.~L.} \bibnamefont{Cho}},
  \bibinfo{author}{\bibfnamefont{Y.~L.} \bibnamefont{Park}},
  \bibinfo{author}{\bibfnamefont{B.~J.} \bibnamefont{Bae}},
  \bibinfo{author}{\bibfnamefont{J.~H.} \bibnamefont{Park}},
  \bibinfo{author}{\bibfnamefont{J.~S.} \bibnamefont{Park}},
  \bibinfo{author}{\bibfnamefont{H.~G.} \bibnamefont{An}},
  \bibinfo{author}{\bibfnamefont{J.~S.} \bibnamefont{Bae}},
  \bibinfo{author}{\bibfnamefont{D.~H.} \bibnamefont{Ahn}},
  \bibnamefont{et~al.}, \bibinfo{journal}{Symp. on VLSI Technology} pp.
  \bibinfo{pages}{102--103} (\bibinfo{year}{2007}).

\bibitem[{\citenamefont{COMSOL}(2014)}]{COMSOL}
\bibinfo{author}{\bibnamefont{COMSOL}}, \emph{\bibinfo{title}{version 4.4}}
  (\bibinfo{publisher}{COMSOL}, \bibinfo{address}{Bangalore, Karnataka, India},
  \bibinfo{year}{2014}).

\bibitem[{\citenamefont{Price}(1956)}]{Price}
\bibinfo{author}{\bibfnamefont{P.~J.} \bibnamefont{Price}},
  \bibinfo{journal}{Phys. Rev.} \textbf{\bibinfo{volume}{104}},
  \bibinfo{pages}{1223} (\bibinfo{year}{1956}).

\bibitem[{\citenamefont{Yan et~al.}(2007)\citenamefont{Yan, Zhu, Zhao, and
  Dong}}]{S_GST}
\bibinfo{author}{\bibfnamefont{F.}~\bibnamefont{Yan}},
  \bibinfo{author}{\bibfnamefont{T.~J.} \bibnamefont{Zhu}},
  \bibinfo{author}{\bibfnamefont{X.~B.} \bibnamefont{Zhao}}, \bibnamefont{and}
  \bibinfo{author}{\bibfnamefont{S.~R.} \bibnamefont{Dong}},
  \bibinfo{journal}{Appl. Phys. A} \textbf{\bibinfo{volume}{88}},
  \bibinfo{pages}{425} (\bibinfo{year}{2007}).

\bibitem[{\citenamefont{Wang et~al.}(2008)\citenamefont{Wang, Lee, Lan, Zhu,
  Josh, Wang, Yang, Muto, Tang, Klatsky et~al.}}]{SiGe}
\bibinfo{author}{\bibfnamefont{X.~W.} \bibnamefont{Wang}},
  \bibinfo{author}{\bibfnamefont{H.}~\bibnamefont{Lee}},
  \bibinfo{author}{\bibfnamefont{Y.~C.} \bibnamefont{Lan}},
  \bibinfo{author}{\bibfnamefont{G.~H.} \bibnamefont{Zhu}},
  \bibinfo{author}{\bibfnamefont{G.}~\bibnamefont{Josh}},
  \bibinfo{author}{\bibfnamefont{D.~Z.} \bibnamefont{Wang}},
  \bibinfo{author}{\bibfnamefont{J.}~\bibnamefont{Yang}},
  \bibinfo{author}{\bibfnamefont{A.~J.} \bibnamefont{Muto}},
  \bibinfo{author}{\bibfnamefont{M.~Y.} \bibnamefont{Tang}},
  \bibinfo{author}{\bibfnamefont{J.}~\bibnamefont{Klatsky}},
  \bibnamefont{et~al.}, \bibinfo{journal}{Appl. Phys. Lett.}
  \textbf{\bibinfo{volume}{93}}, \bibinfo{pages}{193121}
  (\bibinfo{year}{2008}).

\bibitem[{\citenamefont{Goldsmid}(2009)}]{goldsmid}
\bibinfo{author}{\bibfnamefont{H.~J.} \bibnamefont{Goldsmid}},
  \emph{\bibinfo{title}{{Introduction to {T}hermoelectricity}}}
  (\bibinfo{publisher}{Springer}, \bibinfo{year}{2009}).

\bibitem[{\citenamefont{Yamada et~al.}(1991)\citenamefont{Yamada, Ohno,
  Nishiuchi, Akahira, and Takao}}]{Yamada}
\bibinfo{author}{\bibfnamefont{N.}~\bibnamefont{Yamada}},
  \bibinfo{author}{\bibfnamefont{E.}~\bibnamefont{Ohno}},
  \bibinfo{author}{\bibfnamefont{K.}~\bibnamefont{Nishiuchi}},
  \bibinfo{author}{\bibfnamefont{N.}~\bibnamefont{Akahira}}, \bibnamefont{and}
  \bibinfo{author}{\bibfnamefont{M.}~\bibnamefont{Takao}}, \bibinfo{journal}{J.
  Appl. Phys.} \textbf{\bibinfo{volume}{69}}, \bibinfo{pages}{2849}
  (\bibinfo{year}{1991}).

\bibitem[{\citenamefont{Pirovano et~al.}(2004)\citenamefont{Pirovano, Lacaita,
  Pellizzer, Kostylev, Benvenuti, and Bez}}]{Reliability_2004}
\bibinfo{author}{\bibfnamefont{A.}~\bibnamefont{Pirovano}},
  \bibinfo{author}{\bibfnamefont{A.~L.} \bibnamefont{Lacaita}},
  \bibinfo{author}{\bibfnamefont{F.}~\bibnamefont{Pellizzer}},
  \bibinfo{author}{\bibfnamefont{S.~A.} \bibnamefont{Kostylev}},
  \bibinfo{author}{\bibfnamefont{A.}~\bibnamefont{Benvenuti}},
  \bibnamefont{and} \bibinfo{author}{\bibfnamefont{R.}~\bibnamefont{Bez}},
  \bibinfo{journal}{IEEE Trans. Electr. Dev.,} \textbf{\bibinfo{volume}{51}},
  \bibinfo{pages}{714–719} (\bibinfo{year}{2004}).

\bibitem[{\citenamefont{Redaelli et~al.}(2008)\citenamefont{Redaelli, Pirovano,
  Locatell, and Pellizzer}}]{Reliability_2008}
\bibinfo{author}{\bibfnamefont{A.}~\bibnamefont{Redaelli}},
  \bibinfo{author}{\bibfnamefont{A.}~\bibnamefont{Pirovano}},
  \bibinfo{author}{\bibfnamefont{A.}~\bibnamefont{Locatell}}, \bibnamefont{and}
  \bibinfo{author}{\bibfnamefont{F.}~\bibnamefont{Pellizzer}},
  \bibinfo{journal}{Non-Volatile Semiconductor Memory Workshop} pp.
  \bibinfo{pages}{39--42} (\bibinfo{year}{2008}).

\bibitem[{\citenamefont{Joseph et~al.}(2008)\citenamefont{Joseph, Happ, Chen,
  Raoux, Chen, Breitwisch, Schrott, Zaidi, Dasaka, Yee et~al.}}]{Etch}
\bibinfo{author}{\bibfnamefont{E.~A.} \bibnamefont{Joseph}},
  \bibinfo{author}{\bibfnamefont{T.~D.} \bibnamefont{Happ}},
  \bibinfo{author}{\bibfnamefont{S.~H.} \bibnamefont{Chen}},
  \bibinfo{author}{\bibfnamefont{S.}~\bibnamefont{Raoux}},
  \bibinfo{author}{\bibfnamefont{C.~F.} \bibnamefont{Chen}},
  \bibinfo{author}{\bibfnamefont{M.}~\bibnamefont{Breitwisch}},
  \bibinfo{author}{\bibfnamefont{A.~G.} \bibnamefont{Schrott}},
  \bibinfo{author}{\bibfnamefont{S.}~\bibnamefont{Zaidi}},
  \bibinfo{author}{\bibfnamefont{R.}~\bibnamefont{Dasaka}},
  \bibinfo{author}{\bibfnamefont{B.}~\bibnamefont{Yee}}, \bibnamefont{et~al.},
  \bibinfo{journal}{Symp. on VLSI Technology} pp. \bibinfo{pages}{142--143}
  (\bibinfo{year}{2008}).

\bibitem[{\citenamefont{Reifenberg et~al.}(2008)\citenamefont{Reifenberg,
  Kencke, and Goodson}}]{Reifenberg}
\bibinfo{author}{\bibfnamefont{J.~P.} \bibnamefont{Reifenberg}},
  \bibinfo{author}{\bibfnamefont{D.~L.} \bibnamefont{Kencke}},
  \bibnamefont{and} \bibinfo{author}{\bibfnamefont{K.~E.}
  \bibnamefont{Goodson}}, \bibinfo{journal}{IEEE Elec. Dev. Letters}
  \textbf{\bibinfo{volume}{29}}, \bibinfo{pages}{1112} (\bibinfo{year}{2008}).

\bibitem[{\citenamefont{Lake and Datta}(1992)}]{Lake}
\bibinfo{author}{\bibfnamefont{R.}~\bibnamefont{Lake}} \bibnamefont{and}
  \bibinfo{author}{\bibfnamefont{S.}~\bibnamefont{Datta}},
  \bibinfo{journal}{Phys. Rev. B} \textbf{\bibinfo{volume}{46}},
  \bibinfo{pages}{4757} (\bibinfo{year}{1992}).

\bibitem[{\citenamefont{Agarwal and Muralidharan}(2014)}]{Akshay}
\bibinfo{author}{\bibfnamefont{A.}~\bibnamefont{Agarwal}} \bibnamefont{and}
  \bibinfo{author}{\bibfnamefont{B.}~\bibnamefont{Muralidharan}},
  \bibinfo{journal}{Applied Physics Letters} \textbf{\bibinfo{volume}{105}},
  \bibinfo{eid}{013104} (\bibinfo{year}{2014}).

\end{thebibliography}
\end{document}